\documentclass[aip,twocolumn]{revtex4}
\usepackage{amssymb}
\usepackage{amsmath}
\usepackage{graphicx}

\begin{document}

\title[Predicting Band Gaps with Hybrid Density Functionals]
  {Predicting Band Gaps with Hybrid Density Functionals}
\author{Alejandro J. Garza}
\affiliation{Department of Chemistry, Rice University,
  Houston, Texas 77251-1892, USA}
\altaffiliation{Present address: The Joint Center for Artificial
  Photosynthesis, Lawrence Berkeley National Laboratory, Berkeley,
  California 94720, USA}
\author{Gustavo E. Scuseria}
\affiliation{Department of Chemistry and Department of Physics and Astronomy, Rice University,
  Houston, Texas 77251-1892, USA}
\email{guscus@rice.edu}

\begin{abstract}
We compare the ability of four popular hybrid density functionals
(B3LYP, B3PW91, HSE, and PBE0) for predicting band gaps
of semiconductors and insulators over a large benchmark set
using a consistent methodology. We observe no significant statistical
difference in their overall performance although the screened
hybrid HSE is more accurate for typical semiconductors.
HSE can improve its accuracy for  large large band gap materials
--without affecting that of semiconductors-- by including a larger
portion of Hartree--Fock exchange in its short range.
Given that screened hybrids are computationally much less expensive
than their global counterparts, we conclude that they are a
better option for the black box prediction of band gaps.
\end{abstract}

\pacs{0000-1111-222}
\maketitle

\textbf{Introduction.}
Band structure calculations are an important application of
electronic structure methods in materials science.
Due to the cost of computing electronic properties for
extended solids, density functional theory (DFT) methods are
most often used for such calculations.
However, local and semilocal density functional
approximations---the most affordable type of Kohn--Sham functionals
for solids---badly underestimate the band gaps of semiconductors
(the materials of principal interest in practical applications)
and insulators due to self-interaction error~\cite{Perdew1983,Mori-Sanchez2008}.
Hybrid functionals that incorporate a fraction of nonlocal Hartree--Fock (HF)
exchange overcome this issue; however, computing
HF exchange in solids is considerably more expensive than evaluating a
semilocal density functional.
Methods based on the GW approximation can also be used to
compute band gaps more accurately, but these techniques are
even more expensive than hybrids.
A good compromise between cost and accuracy is provided by short-range
screened hybrids: functionals that include HF exchange only
for the short-range part of the electron--electron interaction,
which significantly reduces the cost of evaluating the nonlocal
HF part of the exchange as compared to standard hybrids~\cite{Heyd2003}.
This type of functionals have
been shown to provide reasonably accurate band gaps for semiconductors,
and variants of the Heyd--Scuseria--Ernzerhof~\cite{Heyd2003,Heyd2006,Henderson2007,Moussa2012,Lucero2012}
(HSE) short-range hybrid
have been widely used for the calculation of semiconductor
band gaps for many years~\cite{Henderson2011}.

A recent study~\cite{Crowley2016} declares the (standard) hybrid
B3PW91 as the winner for resolving the band gap prediction problem
for materials design. However, that study did not evaluate the
performance of other hybrid functionals using a direct
comparison based on a consistent methodology. We here carry out
such detailed comparison. We also comment upon the pitfalls
that arise when benchmarking density functionals~\cite{Civalleri2012,Savin2014},
as well as aspects of computational efficiency for solids
between short-range hybrids like HSE~\cite{Heyd2004} and standard global hybrids
(by global, we mean standard full-range hybrids like B3PW91).

The present study focuses on band gap prediction by both short-range screened and global hybrids.
We benchmark popular hybrids like HSE06 (henceforth referred to as simply HSE),
B3PW91, B3LYP, and PBE0 (also known as PBEh or PBE1PBE).
For this comparison, we utilize a variety of error measures in order to
avoid the difficulties and ambiguities that emerge when trying to
decide which electronic structure method is best for a determined task~\cite{Civalleri2012,Savin2014}.
We find that these four functionals are similarly adequate for the
calculation of band gaps; different error measures for
HSE, B3PW91, and B3LYP are very close to each other and,
while PBE0 tends to overestimate band gaps, the error is systematic and
straightforward to correct with a linear fit. If the benchmark set is
narrowed to traditional semiconductors then HSE emerges as the winner.
In addition, we show that short-range hybrids can describe even
large band gap insulators by including an increased portion of
HF exchange in the short range only. Due to their lower cost as compared to
global hybrids, greater ease to achieve convergence, and overall good performance,
we conclude that screened hybrids are a better option for
the prediction of band gaps.

\begin{figure}
  \centering
  \includegraphics[width=0.49\textwidth]{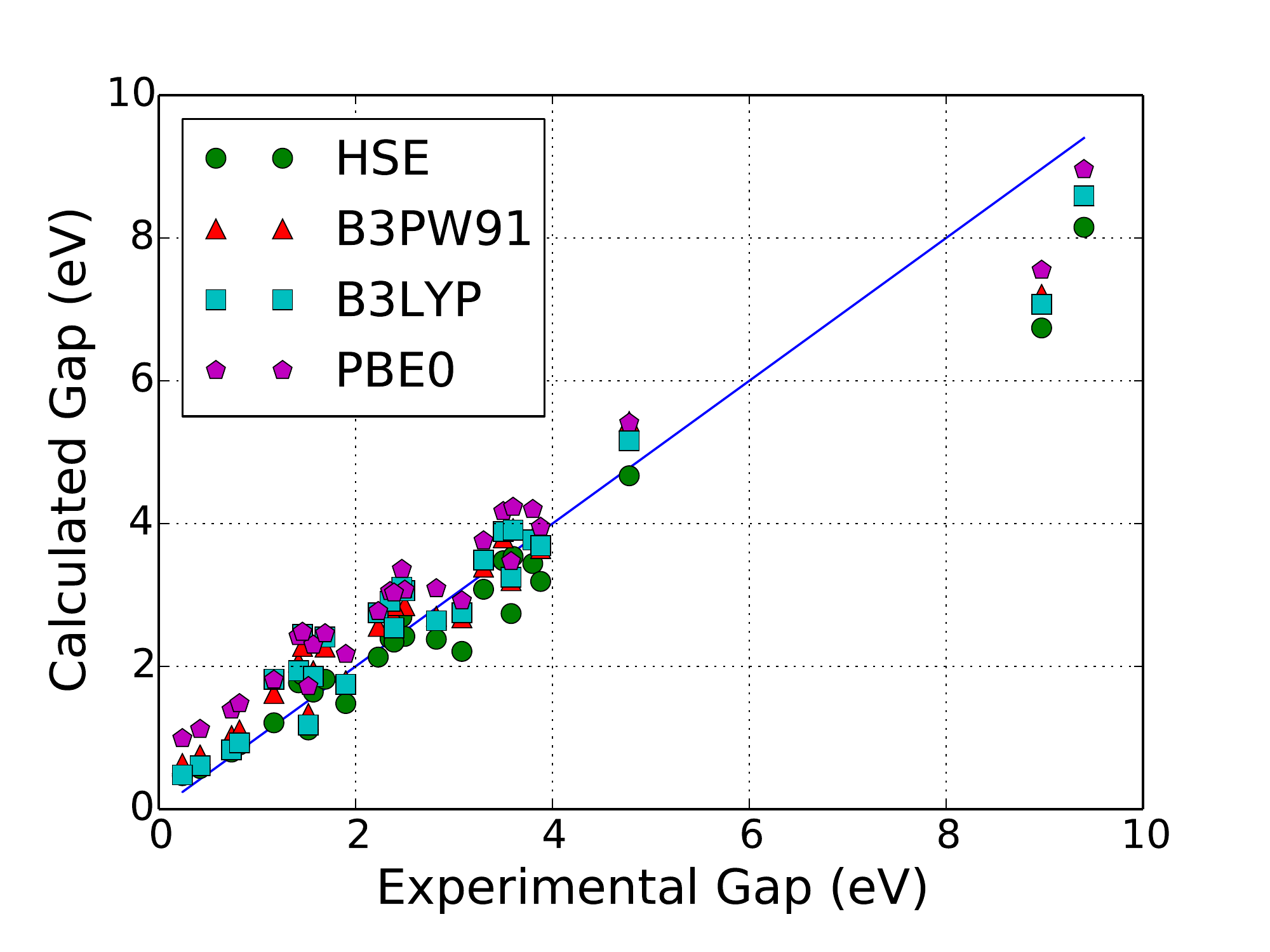}
  \caption{Calculated \textit{vs} experimental  band gaps for the functionals
    studied here.
 }
  \label{fig:1}
\end{figure}

\textbf{Statistical Evaluation.}
We have carried out band gap calculations for 41 semiconductors and
insulators using the \textsc{Gaussian}~\cite{Gaussian09} suite of programs.
The set includes materials from the SC40 dataset~\cite{Lucero2012}, as well as
transition metal oxides (FeO, CoO, NiO, MnO, and VO$_2$) and
large band gap salts (NaCl, LiCl, and LiF); the set is similar
to that of Ref.~\citenum{Crowley2016}
(some compounds have been excluded because
spin-orbit effects are very large, and we have not used spin-orbit corrections in our calculations).
As in that reference, we employed
experimental geometries. The basis sets used for SC40 compounds are the same
as those of Ref.~\citenum{Lucero2012} (which are similar or identical to those in Ref.~\citenum{Crowley2016})
and the basis sets of Ref.~\citenum{Crowley2016} are used  for the rest of the compounds.
However, while we had no problems to compute band gaps
for all these compounds with HSE, some of the global
hybrid calculations (particularly PBE0) were too expensive
or too difficult to converge, even when starting from a converged HSE guess.
Inaccuracies in building an approximate Fock matrix for full-range hybrids make these calculations 
harder to converge, specially for compounds with smaller gaps~\cite{Natiello1984}. 
We were thus unable to obtain results for all 41 compounds with global hybrids.
The calculated \textit{vs} experimental band gaps
for the 27 compounds for which all four functionals successfully
converged are shown in Figure~\ref{fig:1} (see detailed data in the Supporting Information).
Various statistics of the error (which we shall discuss in short and are a compilation of those
presented in Ref.~\citenum{Civalleri2012})
for these calculations and for all of the compounds for which each
functional converged are summarized in Table~\ref{tab:1}.
As shown by the insignificant changes ($\approx$ 0.01--0.03 eV)
in the different error measures
when increasing the size of the benchmark set---from 27 to 41 for HSE, 37 for B3PW91,
and 34 for B3LYP---, the 27 seven compounds for which all four functionals converged
are representative enough to get accurate estimates of the expected deviations from
experiment for each functional.

The first thing that is noticeable from Figure~\ref{fig:1} is that the four
functionals follow the same, fairly linear, trend.
Results from HSE, B3LYP, and B3PW91 are close together; the latter two in particular
give extremely similar band gaps. This is most likely due to the incorporation of
similar amounts of HF exchange: B3LYP and B3PW91 have both 20\% full-range
nonlocal exchange, while HSE includes 25\% but only in the short range.
Inclusion of larger fractions of exact exchange increases the calculated
band gaps:
PBE0 includes 25\% full-range HF exchange, which leads to an overestimation
of the band gaps as compared to experiment and the other functionals;
however, all four functionals underestimate very large insulator band gaps and follow
the linear trend mentioned above.
In fact, if one fits the data in Figure~\ref{fig:1} to a linear function, the
slopes for the four hybrids differ by no more than about 3\% (see Table~\ref{tab:1}).
HSE, however has a linear fit intercept much smaller than the global hybrids (0.36 eV, as compared
to 0.65, 0.63, and 0.95 eV for B3PW91, B3LYP, and PBE0, respectively),
indicating better agreement with the experimental data.

\begin{table*}
\caption{Different band gap prediction error measures for the four functionals
studied here.}
\label{tab:1}
\scalebox{1}{
\begin{tabular}{@{\extracolsep{4pt}}lcccccccc@{}}
  \hline
   & Expt.  & \multicolumn{2}{c}{HSE} & \multicolumn{2}{c}{B3PW91} & \multicolumn{2}{c}{B3LYP} & PBE0 \\
  \cline{3-4}  \cline{5-6} \cline{7-8}
  Size of Test Set &          & 27 & 41 & 27 &  37  & 27 & 34 & 27 \\
  \hline
  Mean Error   & 0 & $-0.24$ & $-0.21$ & $0.14$ & 0.12 & 0.13 & 0.16 & 0.43 \\
  Mean Absolute Error & 0 & 0.37 &  0.38  & 0.44 & 0.41 & 0.44 & 0.47 & 0.59 \\
  Mean Absolute Percent Error & 0 & 16 & 15 & 27 & 26 & 23 & 25 & 45 \\
  Standard Deviation & 0 & 0.55 &  0.54   & 0.53 & 0.51 & 0.56 & 0.56 & 0.50 \\
  Median & 0 & $-0.06$ &   $-0.07$  & 0.33 & 0.29 & 0.19 & 0.27 & 0.63 \\
  Median Absolute Deviation & 0 & 0.21 & 0.27  & 0.25 & 0.29 & 0.34 & 0.3 & 0.14 \\
  Interquartile Range & 0 & 0.49 &  0.50   & 0.57 & 0.55 & 0.69 & 0.73 & 0.43 \\
  Linear Fit Slope & 1 & 0.79 & 0.87   & 0.82 & 0.82 & 0.82 & 0.81 & 0.82 \\
  Linear Fit Intercept & 0 & 0.36 & 0.23   & 0.65 & 0.65 & 0.63 & 0.70 & 0.95 \\
  Kendall $\tau$ Correlation & 1 & 0.85  & 0.88  & 0.85 & 0.88 & 0.84 & 0.85 & 0.86 \\
  \hline
\end{tabular}%
}
\end{table*}

We now discuss the different error measures in Table~\ref{tab:1}.
The first two, the mean error (ME) and the mean absolute error (MAE) are some of
the most widely used error measures in quantum chemistry.
The ME shows that HSE has a tendency to underestimate band gaps ($-0.24$ eV),
while B3PW91 and B3LYP slightly overestimate (0.14 and 0.13 eV, respectively), and PBE0
significantly overestimates (0.43 eV).
Based on MAE, HSE predicts band gaps closer to experiment (MAE = 0.37 eV) and is closely
followed by B3PW91 and B3LYP (0.44 eV for both). PBE0 has the worst MAE (0.59 eV) mainly due
to the aforementioned tendency to overestimate band gaps.
In terms of the mean absolute percent error (MAPE), HSE also comes out better (16\%)
than the other hybrids (27, 23, and 45\% for B3PW91, B3LYP, and PBE0 respectively).
The lower MAPE of HSE as compared to the other hybrids is more notable than its
lower MAE, in part because
the average error for HSE is increased substantially by the errors for the high band gap compounds:
if the very large band gap (\textit{i.e.}, more than $\approx$ 8 eV)
compounds are excluded, the MAEs for HSE, B3PW91, B3LYP, and PBE0
are 0.26, 0.37, 0.36, and 0.57 eV, respectively.
Furthermore, if we consider only typical semiconductors for which low-temperature experimental
band gaps are available (group 1 of the SC40 set; see Ref.~\citenum{Lucero2012}),
the MAEs (in the same order) are 0.18, 0.32, 0.31, and 0.55 eV, and
the MAPEs are 17, 33, 26, and 62.
We also report in Table~\ref{tab:1} the standard deviations (SDs) of the error; a perfect
approximation would have zero ME and zero SD. However, no significant difference is observed between
the SDs of the hybrids: the largest difference is 0.05 eV between HSE and PBE0. Note also that
the functional with the worst ME and MAE, PBE0, has the smallest SD.
Thus, as has been pointed out by Savin \textit{et al.}~\cite{Civalleri2012,Savin2014},
using different error measures to judge
the quality of a functional can lead to contradictory results.

The following three statistics of the deviations from experiment in Table~\ref{tab:1},
the median, the median absolute deviation, and
the interquartile range, are measures taken from robust statistics,
which are more resilient to outliers as compared to the ME, MAE, and SD.
In statistics, the median is considered a better descriptor of typical outcome
than the mean when the latter is biased by outliers;
the median absolute deviation measures statistical dispersion
(in this case of the error); and the interquartile range measures the
variability in a set of data. For a method that perfectly predicts band gaps,
all these values must be zero.
We observe,
as before, that different functionals provide the best results for different error measures:
HSE has the best median and PBE0 the worst, but the latter has the best median absolute deviation
and interquartile range.
Note, however, that the median absolute deviation and the
interquartile range of HSE is only 0.06 eV larger than that of PBE0, while
the median of HSE is 0.53 eV closer to zero than the median of PBE0.
Thus, overall, measures from robust statistics favor HSE the most.

Apart from the various statistics analyzed so far, an important feature
of a quantum chemical method is its ability to correctly reproduce trends.
A trend can be defined by a linear fit as follows:
\begin{equation}
  y_\text{expt} = m y_\text{calc} + b.
\end{equation}
Table~\ref{tab:1} shows the values of the slope $m$ and the intercept $b$
from this linear fit for the band gaps calculated by the four hybrid
functionals.
Ideally, $m = 1$ and $b = 0$.
Judging by the values of the slopes, the four hybrids have a similar quality
for predicting trends: the slopes are all around 0.80 and do not differ by more than about 3\%.
However, PBE0 has a substantially higher intercept (0.95 eV) than other methods, whereas
HSE is closest to zero (0.36 eV), and B3PW91 and B3LYP are in between that ($\approx$ 0.65 eV).
Hence, with respect to trends, all functionals are similar but HSE appears to
be slightly better again.

Lastly, let us consider a measure of rank correlation to judge the quality
of predicted band gaps: the Kendall $\tau$ rank correlation coefficient.
In order to construct this coefficient, we first need to define the pairs
$(g_i^\text{expt},g_i^\text{calc})$, where $g_i^\text{expt}$ and $g_i^\text{calc}$
are the experimental and calculated gaps, respectively.
The pairs $i$ and $j$ and then said to be concordant if $g_i^\text{expt} < g_j^\text{expt}$
and $g_i^\text{calc} < g_j^\text{calc}$, or if $g_i^\text{expt} > g_j^\text{expt}$ and
$g_i^\text{calc} > g_j^\text{calc}$; tied if $g_i = g_j$; and
discordant otherwise. The Kendall $\tau$ correlation coefficient is then
\begin{equation}
  \tau = \frac{P - Q}{\sqrt{(P+Q+T)(P+Q+U)}}
\end{equation}
where $P$ is the number of concordant pairs, $Q$ the number of discordant pairs, and
$T$ and $U$ the number of ties in the experimental and calculated set of band gaps, respectively.
Thus, $\tau$ measures the similarity in the orderings of two datasets: a $\tau$ value of 1 indicates
perfect agreement in the ordering, whereas $\tau = -1$ indicates complete disagreement.
It can, in a way, be considered a measure of qualitative agreement.
We see in Table~\ref{tab:1} that all four hybrids provide essentially the same $\tau$ value
($\approx $0.85), and thus have all similarly good qualitative agreement with experiment.

\begin{table}
  \caption{Wilcoxon signed-rank test $W$ values for the absolute errors of pairs of hybrid functionals
    using the data that appears in Figure~\ref{fig:1}.
  The associated $p$-values  are included in parenthesis. The value for $W_\text{crit}$ with a
  95\% confidence interval is 107; $W < W_\text{crit}$ for a pair of methods indicates 
  that their absolute errors show the same statistical distribution around zero. 
  }
\label{tab:2}
\scalebox{0.9}{
\begin{tabular}{@{\extracolsep{4pt}}lcccc@{}}
  \hline
   & HSE & B3PW91 & B3LYP & PBE0 \\
  \hline
  HSE & -- & 42(0.61) & 84(0.31) & 128(0.14) \\
  B3PW91 & 42(0.61) & -- & 18(0.83) & 227(0.01) \\
  B3LYP & 84(0.31) & 18(0.83) & -- & 214(0.01) \\
  PBE0 & 128(0.14) & 227(0.01) & 214(0.01) & -- \\
  \hline
\end{tabular}%
}
\end{table}

One could also analyze the distribution of the error of pairs of methods, 
and ask whether the distributions differ statistically. 
This is done in Table~\ref{tab:2} using the Wilcoxon signed-rank test 
for the absolute error data.
The null hypothesis of the test is that the difference between pairs (of absoulte errors for two methods)
follows a symmetric distribution around zero because each set has the same statistical distribution, 
resulting in similar positive and negative differences;
the null hypothesis is rejected when the test statistic $W$ is greater than a certain 
$W_\text{crit}$ determined by the sample size and level of confidence. 
A $p$-value associated with the test represents the probability of 
obtaining the observed distribution, or a more asymmetric distribution around zero, 
assuming that the null hypothesis is true; $p$-values larger than 0.05 
have been related with a lack of meaningful difference in the context of 
method comparison~\cite{Skone2016}. 
Based on both the $W$ test statistic and its associated $p$-values,
the results in Table~\ref{tab:2} indicate no significant statistical difference 
(considering a commonly used 95\% confidence interval)
in the absolute error distributions of HSE, B3PW91, and B3LYP. 
The $W$ statistics suggest, however, that differences exist between 
these three functionals and PBE0; the $p$-values also imply significant
difference between PBE0 and both B3PW91 and B3LYP. 
If the same analysis is carried out for signed error and absolute 
percentage error, all pairs of functionals come out as being statistically different 
except for the B3PW91/B3LYP pair. 
The results of the Wilcoxon signed-rank test support the more qualitative 
comparison of the ME, MAE, and MAPE discussed above, and put such a discussion 
in a rigorous statistical framework.

\textbf{On the underestimation of large band gaps.}
As shown here and in other works~\cite{Civalleri2012},
even hybrid functionals underestimate band gaps when these are
considerably large. No single hybrid with a fixed amount of HF exchange can accurately
describe small and large band gap materials simultaneously: the fraction of nonlocal exchange needed in a
solid depends on its dielectric function~\cite{Skone2016,Marques2011,Skone2014}.
Increasing the percentage of HF exchange in a hybrid functional increases the band gap, which allows
for the description of larger band gap insulators, but causes overestimation of semiconductor band gaps.
A way to optimize the fraction of HF exchange in hybrid functionals based on a local estimator of
the dielectric function has been proposed in Ref.~\citenum{Marques2011}.
With this procedure, one obtains, for example, that HSE predicts a band gap 22.29 eV for solid Ne
(which has an experimental gap of 21.70 eV), while still giving good semiconductor band gaps
(\textit{e.g.}, 0.82 eV for Ge, compared to 0.74 eV from experiment)~\cite{Marques2011}.
The method is applicable to global hybrids too: an optimized PBE0 gives gaps of 21.88 and 0.68 eV for
Ne and Ge, respectively, compared to 15.14 and 1.31 eV predicted by standard PBE0~\cite{Marques2011}.

\begin{figure}
  \centering
  \includegraphics[width=0.49\textwidth]{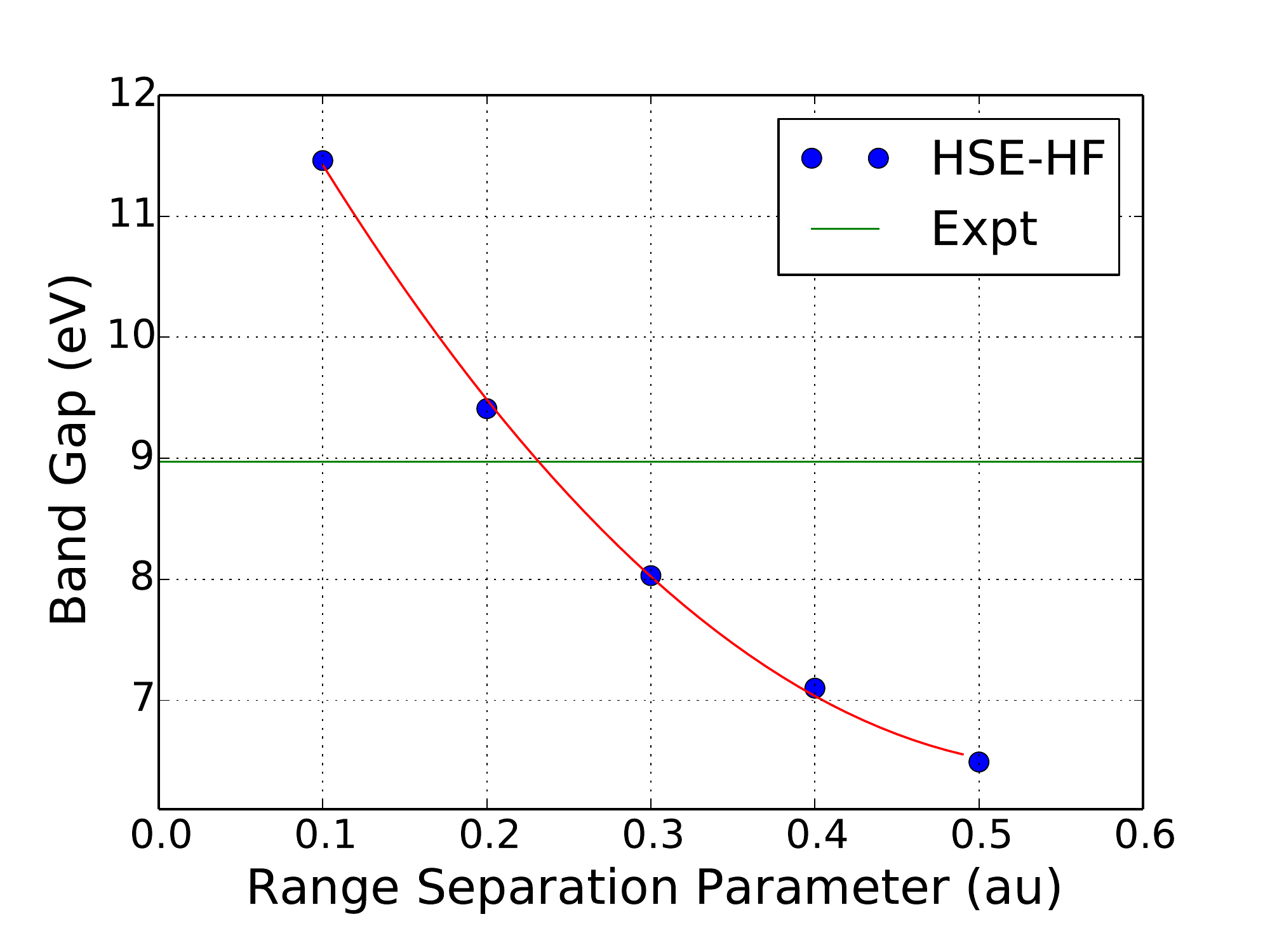}
  \caption{NaCl band gap calculated by HSE-HF (HSE06 wtih 100\% exact exchange) as a
    function of the range-separation parameter.
 }
  \label{fig:2}
\end{figure}

As another example of how even screened hybrids can describe large band gap insulators,
consider the HSE-HF functional of Ref.~\citenum{Garza2014}, which is simply HSE06 but with 100\%
exact exchange in the short range.
Figure~\ref{fig:2} shows the band gap of NaCl calculated by HSE-HF as a function of the range
separation parameter, $\omega$.
The smaller this parameter is, the larger the fraction of HF exchange incorporated.
At small $\omega$ values, the band gap is overestimated; at $\omega \approx 0.2$--$0.3$ au the
value is close to experiment; and at higher $\omega$ values the gap is underestimated.
This is just one example, but this behavior is very general and this observation is in agreement
with previous works showing that long-range HF exchange is seldom necessary and that the physics
behind the more realistic properties computed with hybrid functionals is the reduction of
the self-interaction error by incorporation of substantial amounts of exact
exchange~\cite{Garza2014,Henderson2009}.
Thus, a short-range screened hybrid can describe even large band gaps insulators and one does not
need to resort to more expensive global hybrids.
Note also that 
although, in principle, long-range HF exchange should be 
included when describing solids, its effect should be mostly canceled by correlation.
Inadequate treatment of correlation
is the reason why the HF method and typical global hybrid functionals cannot describe
metallic behavior.
Hence, it is convenient to eliminate long-range exchange and use
a screened hybrid to more adequately describe the physics of an
extended solid.

\textbf{Additional Considerations.}
We have focused on band gaps calculated using experimental geometries
and observed little overall difference between the different hybrids tested.
However, other properties, including geometries, can be significantly
different. For example, the MAE of B3LYP for lattice constants
(0.053 \AA) is more than twice larger than the MAEs of HSE and
PBE0 (0.024 and 0.022 \AA, respectively)~\cite{Paier2007}. Atomization energies
of solids are also significantly worse when
calculated with B3LYP (MAE = 0.59 eV) as compared to
HSE and PBE0 (MAE $\approx$ 0.26 eV); B3PW91
improves over B3LYP, but it is not as good as the latter two
functionals (MAE = 0.39 eV)~\cite{Paier2007}. 
The failure of B3LYP is attributed to not correctly reproducing the 
homogeneous electron gas behavior, and B3PW91 improves upon it by 
fulfilling this limit~\cite{Paier2007}.

Even when considering only band gaps and using experimental
geometries, another important consideration arises: the quality of the
experimental data. In particular for the band gaps, there can be a
significant variability depending on factors such as the method of
measurement and temperature (which sometimes is not reported).
For example, band gap values of 3.6, 2.6, 2.5, 2.1, 2.7, and 5.43 eV have been
reported for CoO (see Supporting Information of Ref.~\citenum{Crowley2016}).
This variability leads to the question: What value should
one use as reference? One can average of all these values (3.16 eV)
and obtain very good agreement with B3PW91 (calculated here as 3.06 eV)~\cite{Crowley2016}.
However, this approach may not be appropriate in this case:
the experimentally reported 5.43 eV value is most likely the direct gap,
whereas the smaller values correspond to the indirect gap~\cite{Gillen2013}.
Even if one decided to follow this methodology, the standard deviation of
the six experimental band gap values for CoO is 1.22 eV, and hence
all functionals predicting values in the wide range between 1.90 and 4.34 eV would have
to be considered to be in agreement with experiment.
The often cited value of 2.5 eV for the CoO band gap is in reasonable agreement
with the HSE indirect gap of 2.82 eV.
An additional consequence of the variability in the reference data is that small
differences in the average errors of different functionals are not really meaningful.
Thus, based on, say, the MAEs in Table~\ref{tab:1}, it would be hard to tell whether
HSE, B3PW91, or B3LYP are truly closer to experiment.
Furthermore, adding to the level of uncertainty,
experimental reference band gaps are assumed to be an ideal measurement
of the fundamental band gap (\textit{i.e.}, the difference between the ionization
potential and electron affinity). However, this
is not always the case~\cite{Civalleri2012}, whereas the
calculation of generalized Kohn--Sham band gaps (as done here) is
a rigorous and honest approximation to the fundamental band gap~\cite{Perdew2016}.
Considering all of this, and the fact that the use of different, finite basis sets leads
to further changes in predicted band gaps, it would be unreasonable to
assert that a certain functional is decidedly better than another for band gap predictions
if the differences in the error measures are small (\textit{e.g.},  
less than about 0.1 eV or, depending on the case,
even more than this, as shown by the CoO example above).
Statistical methods such as the ones used here should be considered for 
assessing whether the differences are significant.

The importance of careful inspection of reference data from the literature and the use of a consistent 
methodology in benchmark calculations applies not only to density functionals.
The GW method, for example, is quite sensitive to a number of variables: 
the input (reference) wavefunction, number of
empty bands included, dielectric matrix energy cut-off, etc.
To illustrate this, consider the GW band gaps of ZnO used as reference to compare against 
B3PW91 in the Supporting Information document of Ref.~\citenum{Crowley2016}. 
The experimental band gap is 3.44 eV; the GW gaps in the literature vary from 0.1 to 4.61 eV. 
The worst value, 0.1 eV, was computed using the static  Coulomb  hole  and  screened  exchange 
(COHSEX) approximation, which is generally not recommended for quantitative calculations~\cite{Huser2013}. 
Shih~\textit{et al.}~\cite{Shih2010} report that conventional GW approaches
can provide accurate estimates of the ZnO band 
gap if convergence in the evaluation of the Coulomb-hole self-energy is carefully resolved.
Further examples with large variability (more than 1 eV) in the GW gaps cited in Ref.~\citenum{Crowley2016} 
include Si (0.56--1.91), GaAs (1.09--3.77), SiC (1.8--2.88), AlP (1.88--3.1), CdS (2.11--3.41), 
CoO (2.4--4.78), GaN (2.75--3.82), MnO (2.34--4.39), ZnS (1.52--4.15), NiO (1.74--5.0), 
diamond (5.59--6.99), LiCl (8.75--10.98), and LiF (13.13--16.17) among others. 
It is worth pointing out here that Ref.~\citenum{Crowley2016} concluded that B3PW91 
was superior to other functionals by comparing against GW, assuming that GW is more accurate than 
these (other) functionals. 
However, as shown here, such comparisons are not straightforward and B3PW91 does not appear 
to have any decisive advantage over other commonly used hybrids 
when the direct comparison is done using identical geometries, basis sets, etc.

The effects of spin-orbit coupling should also be mentioned here. 
We have not included these effects in our calculations, as such corrections 
are not widely available in commercial codes. 
Nevertheless, our conclusions are not affected by neglecting spin-orbit 
coupling because of, mainly, two reasons. The first one is that 
we have not included compounds for which spin-orbit coupling effects are 
important in our test set. 
The second one is related to the fact that spin-orbit corrections 
apply in the same way to all functionals:
\textit{i.e.}, if we use experimental spin-orbit coupling values to correct the band gaps, 
the shift in the error is the same for every functional.
Here, we observe that 
various hybrids provide rather similar results, with no clear 
winner in all cases and a different method being ``better'' depending on the 
error measure. After spin-orbit coupling corrections, the functionals will still 
provide similar results because these corrections are identical for all of them.

Lastly, an important consideration for calculations on extended systems is 
computational cost.
The increase in CPU time when going from a semilocal functional, 
to a screened hybrid, to a global hybrid depend on the type of basis set employed. 
Here, we have used codes based on localized gaussian basis sets, which are not as efficient 
for semilocal functionals but have a less dramatic increase in cost for hybrids
as compared to plane-wave based codes. This means that,
in plane-wave implementations of DFT, 
screened hybrids are more strongly favored over global hybrids concerning CPU time.

\textbf{Conclusions.}
After carrying out a thorough comparison between four popular hybrid functionals
for predicting band gaps using a consistent methodology in the calculations,
we did not find truly significant differences in the overall performance of three of them:
HSE, B3PW91, and B3LYP (the latter two in particular give very similar gaps).
PBE0 tends to overestimate band gaps but the error is systematic and the statistical analysis
shows that all four functionals provide good qualitative agreement with the reference data.
Nevertheless, our results do suggest that HSE is more accurate for typical semiconductors.
The analysis here shows that there is nothing particularly remarkable
about the band gaps predicted by B3PW91~\cite{Crowley2016} when they are
compared to other global hybrids using a similar amount of HF exchange.
We have also shown how screened hybrids can incorporate larger fractions
of HF exchange in the short range to compensate for the effects
that result from screening in large gap insulators, 
and ways to tune the mixing parameter that solve this 
problem have also been proposed in the literature~\cite{Skone2016,Marques2011,Skone2014}.
Global hybrids also underestimate very large band gaps, and can too be modified to 
include more exchange to solve this problem, but at a higher computational cost. 
Thus, based on cost and accuracy considerations,
we recommend HSE for semiconductors and tuned versions of HSE for larger gap cases.

In summary, screened hybrids can describe the same type of problems
that global hybrids do, but in addition, they provide access to modelling metallic behavior,
have a much lower computational cost, and are easier to converge.
Our conclusion is therefore that a screened hybrid like HSE is preferable
over global hybrids for the black box prediction of band gaps of novel materials.


\textbf{Acknowledgements.}
This work was supported as part of the Center for the Computational
Design of Functional Layered Materials, an Energy Frontier Research
Center funded by the US Department of Energy, Office of Science, Basic
Energy Sciences under Award [\# DE-SC0012575].
G.E.S. is a Welch Foundation chair (C-0036). 
We also thank the authors
of Ref~\citenum{Skone2016} for bringing their paper to our attention and suggesting
the use of the Wilcoxon signed-rank test.

\textbf{Supporting Information Available.}
  Details about the calculations, as well as band gap data
  used in this work.


\begin{thebibliography}{}

\bibitem{Perdew1983}
  Perdew, J. P.; Levy, M. Physical Content of the Exact Kohn--Sham
  Orbital Energies: Band Gaps and Derivative
  Discontinuities. \emph{Phys. Rev. Lett.} \textbf{1983}, \emph{51}, 1884--1887.


\bibitem{Mori-Sanchez2008}
  Mori--S\'{a}ńchez P.; Cohen, A. J.; Yang, W. Localization and Delocalization
  Errors in Density Functional Theory and Implications for Band-Gap
  Prediction. \emph{Phys. Rev. Lett.} \textbf{2008}, \emph{100}, 146401.

\bibitem{Heyd2003}
  Heyd, J.; Scuseria, G. E.; Ernzerhof, M.
  Hybrid Functionals Based on a Screened Coulomb Potential.
  \emph{J. Chem. Phys.} \textbf{2003},  \emph{118}, 8207--8215.

\bibitem{Heyd2006}
  Heyd, J.; Scuseria, G. E.; Ernzerhof, M. Erratum: ``Hybrid
  functionals based on a screened Coulomb potential''. \emph{J. Chem. Phys.}
  \textbf{2006}, \emph{124}, 219906.

\bibitem{Henderson2007}
  Henderson, T. M.; Izmaylov, A. F.; Scuseria, G. E.;
  Savin, A. The Importance of Middle-Range Hartree--Fock-type
  Exchange for Hybrid Density Functionals.
  \emph{J. Chem. Phys.} \textbf{2007},  \emph{127},
  221103.

\bibitem{Moussa2012}
  Moussa, J. E.; Schultz, P. A.; Chelikowsky, J. R.
  Analysis of the Heyd-Scuseria-Ernzerhof Density Functional Parameter Space.
  \emph{J. Chem. Phys.} \textbf{2012}, \emph{136}, 204117.

\bibitem{Lucero2012}
  Lucero, M. J.; Henderson, T. M.; Scuseria, G. E. Improved
  Semiconductor Lattice Parameters and Band Gaps from a Middle-Range
  Screened Hybrid Exchange functional.
  \emph{J. Phys.: Condens. Matter} \textbf{2012}, \emph{24}, 145504.


\bibitem{Henderson2011}
  Henderson, T. M.; Paier, J.; Scuseria, G. E.
  Accurate treatment of solids with the HSE Screened Hybrid.
  \emph{Phys. Status Solidi B} \textbf{2011}, \emph{248}, 767--774.

\bibitem{Crowley2016}
  Crowley, J. M.; Tahir-Kheli, J.; Goddard III, W. A.
  Resolution of the Band Gap Prediction Problem for Materials Design.
  \emph{J. Chem. Phys. Lett.} \textbf{2016}, \emph{7}, 1198--1203.

\bibitem{Civalleri2012}
  Civalleri, B.; Presti, D.; Dovesi, R.; Savin, A.
  On Choosing the Best Density Functional Approximation.
  \emph{Chem. Modell.} \textbf{2012}, \emph{9}, 168--185.

\bibitem{Savin2014}
  Savin, A.; Johnson, E.
  Judging Density-Functional Approximations: Some Pitfalls of Statistics.
  \emph{Top. Curr. Chem.} \textbf{2015}, \emph{365}, 81--96.

\bibitem{Heyd2004}
  Heyd, J. Screened Coulomb Hybrid Density Functionals.
  Ph.D. Dissertation, Rice University, Houston, TX, 2004.

\bibitem{Gaussian09}
    Frisch, M. J.; Trucks, G. W.; Schlegel, H. B.;
    Scuseria, G. E.; Robb, M. A.; Cheeseman, J. R.;
    Scalmani, G.; Barone, V.; Mennucci, B.;
    Petersson, G. A.  et al.
    \emph{Gaussian 09}, Revision A.02;
    Gaussian Inc.: Wallingford, CT, 2009.

\bibitem{Natiello1984}
  Natiello, M. A.; Scuseria, G. E. 
  Convergence properties of Hartree--Fock SCF Molecular Calculations.
  \emph{Int. J. Quantum Chem.} \textbf{1984}, \emph{26}, 1039. 

\bibitem{Skone2016}
  Skone, J. H.; Govoni, M.; Galli, G. 
  Nonempirical Range-Separated Hybrid Functionals for Solids and Molecules.
  \emph{Phys. Rev. B} \textbf{2016}, \emph{93}, 235106.  

\bibitem{Marques2011}
  Marques, M. A. L.;  Vidal, J.;  Oliveira, M. J. T.;   Reining, L.;   Botti, S.
  Density-based Mixing Parameter for Hybrid Functionals,
  \emph{Phys. Rev. B} \textbf{2011}, \emph{83}, 035119.

  \bibitem{Skone2014}
    Skone, J. H.; Govoni, M.; Galli, G. Self-Consistent Hybrid Functional
    for Condensed Systems, \emph{Phys. Rev. B} \textbf{2014}, \emph{89}, 195112.

\bibitem{Garza2014}
  Garza, A. J.; Wazzan, N. A.; Asiri, A. M.; Scuseria, G. E.
  Can Short- and Middle-Range Hybrids Describe the Hyperpolarizabilities of 
  Long-Range Charge-Transfer Compounds?
  \emph{The Journal of Physical Chemistry A} \textbf{2014} \emph{118},
  11787--11796.

\bibitem{Henderson2009}
  Henderson, T. M.; Izmaylov, A. F.; Scalmani, G.;
  Scuseria, G. E.
  Can Short-Range Hybrids Describe Long-Range-Dependent Properties?
  \emph{J. Chem. Phys.} \textbf{2009},  \emph{131}, 044108.

\bibitem{Paier2007}
  Paier, J.;  Marsman, M.;  Kresse, G.
  Why Does the B3LYP Hybrid Functional Fail for Metals?
  \emph{J. Chem. Phys.} \textbf{2007}, \emph{127}, 024103.

\bibitem{Gillen2013}
  Gillen, R.; Robertson, J.
  Accurate Screened Exchange Band Structures for the Transition Metal Monoxides MnO, FeO, CoO and NiO.
  \emph{J. Phys.: Condens. Matter} \textbf{2013}, \emph{25}, 165502.

\bibitem{Perdew2016}
  Perdew, J. P.; Yang, W.; Burke, K.; Yang, Z.; 
  Gross, E. K. U.; Scheffler, M. Scuseria, G. E.;
  Henderson, T. M.; Zhang, I. Y.; Ruzsinszky, A.; 
  Peng, H.; Sun, J.
  Understanding Band Gaps of Solids 
  in Generalized Kohn--Sham Theory. 
  \textbf{2016}, arXiv:1608.06715. arXiv.org e-Print archive. 
  https://arxiv.org/abs/1608.06715 (accessed Oct 2, 2016). 

\bibitem{Huser2013} 
  H\"{u}ser, F; Olsen, T.; Thygesen, K. S. 
  Quasiparticle GW Calculations for Solids, Molecules, and Two-Dimensional Materials. 
  \emph{Phys. Rev. B} \textbf{2013}, \emph{87}, 235132. 

\bibitem{Shih2010}
  Shih, B.-C.; Xue, Y.; Zhang, P.; Cohen, M. L.; Louie, S. G. 
  Quasiparticle Band Gap of ZnO: High Accuracy from the Conventional G$^0$W$^0$
  Approach. 
  \emph{Phys. Rev. Lett.} \textbf{2010}, \emph{105}, 146401. 



\end{thebibliography}
\end{document}